\documentclass[review,3p,12pt]{elsarticle}

\usepackage{amssymb}
\usepackage{amsmath}
\usepackage[ruled,linesnumbered]{algorithm2e}
\usepackage{setspace}
\usepackage{siunitx}
\usepackage{lineno}

\usepackage{multirow}
\usepackage{booktabs}
\usepackage{overpic}
\usepackage[colorlinks,linkcolor=blue,anchorcolor=blue,citecolor=blue]{hyperref}

\journal{Soft Matter}

\begin{document}

\begin{frontmatter}


  \title{\large{How do 3M Command™ strips work? A fracture mechanics approach}}

  \pdfstringdefDisableCommands{%
    \def\corref#1{}%
  }

  \author[cornell]{Xue-Ling Luo\corref{cor1}}
  \ead{xl2254@cornell.edu}
  \author[cornell,pasteur]{Nikolaos Bouklas}
  \author[cornell,hok]{Chung-Yuen Hui\corref{cor1}}
  \ead{ch45@cornell.edu}

    \cortext[cor1]{Corresponding authors.}
  
    \affiliation[cornell]{addressline={Sibley School of Mechanical and Aerospace Engineering, Cornell University},
              city={Ithaca},
              state={NY},
              country={USA}}

    \affiliation[pasteur]{addressline={Pasteur Labs},
              city={Brooklyn},
              state={NY},
              country={USA}}

    \affiliation[hok]{addressline={Soft Matter GI-CoRE, Hokkaido University},
              city={Sapporo},
              country={Japan}}

  \begin{abstract}
    Removable adhesive systems such as 3M Command™ strips are designed to support substantial loads while allowing clean, damage-free removal from the substrate. These systems rely on a highly extensible adhesive strip that bonds strongly during use but releases when stretched, causing the adhesive layer to elongate and progressively debond from the surfaces. A central challenge in the design of stretch-release adhesives is therefore to maximize load-bearing capacity while minimizing the force required for removal. This study investigates the finite-deformation mechanics governing both load support and tape release in a hyperelastic stretch-release adhesive system, with particular focus on the 3M Command™ tape geometry. Explicit analytical expressions are derived for the energy release rate of interfacial cracks under both load-bearing and release conditions and are validated against $J$-integral evaluations from finite element simulations. The results show that the ratio of maximum supported load to release force scales linearly with the ratio of bonded length to adhesive thickness, which is typically very large. We also investigate geometry-driven alternating crack propagation between the backing and substrate interfaces, governing tape removal, by analytical solutions and simulations. Parametric studies of competing interfacial fracture toughnesses produce failure envelopes that provide a predictive framework for estimating release forces and unstable crack propagation in multilayer stretch-release adhesive systems.
  \end{abstract}



\end{frontmatter}


\section{Introduction}

\begin{figure}[t]
    \centering
    \includegraphics[width=0.88\linewidth]{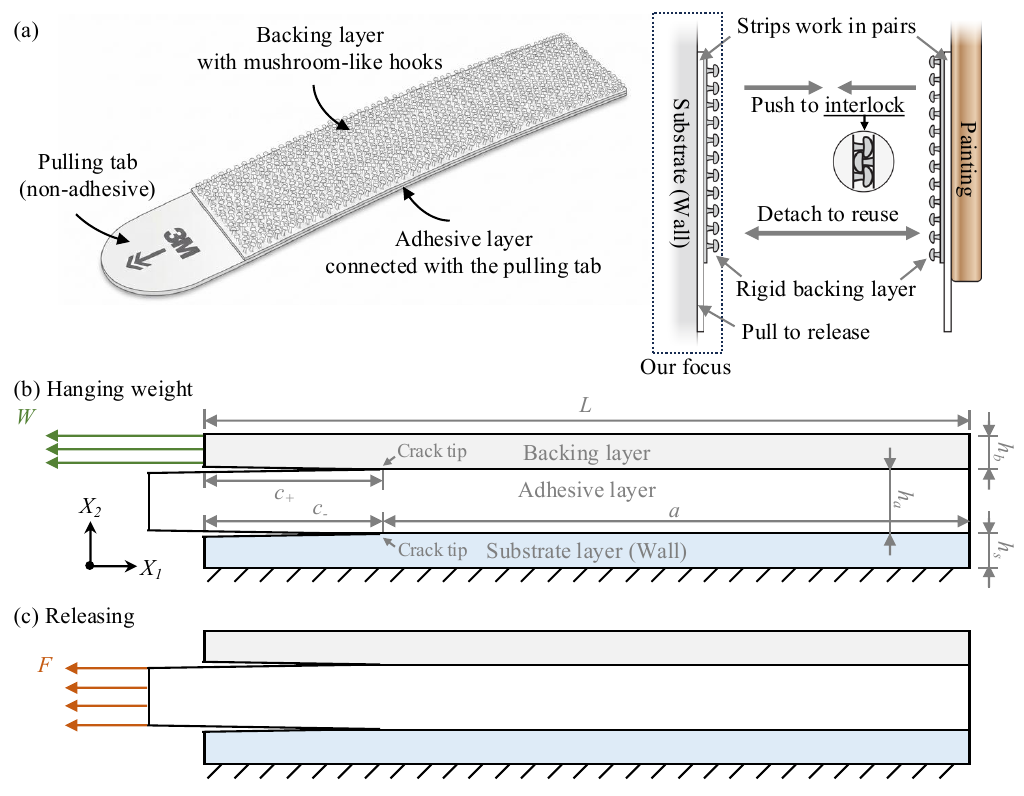}
    \caption{(a) Schematic of the 3M Command™ Picture Hanging strip, how the strips work in pairs, and our focused region. (b)-(c) The schematic geometry of our focused region and boundary conditions under two loading cases. The backing layer in (b) and (c) represents one layer of the interlocking stiff mushroom fibrils together with its thin rigid backing, colored in gray in (a). The bonded portion of the adhesive layer has length $a$ (bond length). The pull-tab consists of the unbonded portion of the adhesive and forms two interface cracks of equal length $c=L-a$, where $L$ is the total length of the tape.  The tape has width $w \gg h_a$ in the out-of-plane direction. The loads applied in the two distinct cases represent (b) a hanging weight and (c) the releasing action.}
    \label{fig:geometry}
\end{figure}

3M Command™ strips typically consist of a multilayer stretch-release adhesive tape bonded between a stiff backing and a “substrate” such as a wall or a glass panel \cite{command_picture_strips_2026} (see Fig.~\ref{fig:geometry}(a) for a schematic of a Command™ Picture Hanging strips). The adhesive strip, of thickness $h_a$, is thin and highly extensible, with one end protruding below the mounted object to form a pull tab. The air gaps among the backing, adhesive, and substrate can be considered as two interface cracks of equal length $c = L-a$, where $L$ is the total length of the Command™ tape (see Fig.~\ref{fig:geometry}(b)-(c)). During normal use of a Command™ Picture Hanging strip attached to the substrate, the adhesive interfaces of length $a$, corresponding to the bond length, sustain shear loading and support the hanging weight $W$. This weight is transmitted by the interlocking mushroom-shaped fibrils through another strip attached to the weight (Fig.~\ref{fig:geometry}(a)). To remove the strip from the substrate, the weight, including the strip attached to it, is removed and the exposed tab is pulled longitudinally with a force $F$ and stretched parallel to the wall, causing the adhesive layer to elongate substantially. As the strip stretches, the interfacial cracks progressively propagate along the adhesive interfaces between the backing layer and the wall, allowing the strip to detach cleanly from the substrate without damaging the wall surface.  A video illustrating the use of this tape can be found on the 3M website \cite{command_picture_strips_2026}. Other 3M Command™ products \cite{command_products_2026} that consist of a single layer of adhesive sticking to both the substrate (a wall) and the weight, such as a Command™ Hook, can also be analyzed in the same way. 

The mechanical stability of the two distinct loading cases, supporting a weight and releasing, depends on the stability of the interface cracks and can be analyzed using a fracture mechanics approach.  The weight-supporting mechanics are similar to a zero-degree peel test.  Kendall \cite{kendall1975thin,kendall1975crack} has established the energy balance for peeling an elastic film with different peel angles on a rigid substrate.  However, his analysis relies on the assumption of linear elasticity and does not capture the finite-deformation hyperelastic response of the film, which governs the mechanics of stretch-release tapes in the zero-degree peel geometry.

To address this limitation, Hui et al. \cite{hui2018mechanics} developed a large-deformation hyperelastic model for the zero-degree peel test using a three-term Yeoh strain-energy function and obtained closed-form solutions for the stress field in the adhesive layer. They showed that at large shear strains, the lateral true stress parallel to the substrate grows quadratically with shear strain, whereas the shear stress grows only linearly, so that the lateral stress dominates, in contrast to Kaelble’s linear theory, which assumes zero lateral stress \cite{kaelble1960theory,hui2018mechanics}. Liu et al. \cite{liu2019mechanics} later extended the analysis to finite bond lengths and derived exact expressions for the energy release rate under large deformation and material nonlinearity. Wang et al. \cite{wang2020strength} further showed that the steady-state energy release rate for a soft elastic layer in shear equals the product of the layer thickness and the strain-energy density, independent of the hyperelastic model. They also identified a characteristic fractoadhesive length above which geometric stress concentrations reduce the measured adhesion strength, and demonstrated through shear-lag analysis that the relative magnitudes of the bonded length and shear-lag length determine whether the adhesive or adherend elasticity controls the energy release rate \cite{wang2021lap}. Finally, finite simple shear can generate normal stresses through the Poynting effect \cite{horgan2017poynting}, and whether this coupling alters the energy release rate depends sensitively on the constitutive form of the strain-energy function.

While the zero-degree peel test studied by Hui et al. \cite{hui2018mechanics} can be modified to study the weight-hanging configuration, the release configuration relevant to 3M Command™ strips (Fig.~\ref{fig:geometry}(c)) remains largely unexplored. In this multilayer adhesive system, cracks may propagate along either the backing–adhesive or adhesive–substrate interface, and the competition between these paths governs the removal mechanics.

In this study, we address this gap by first revisiting classical analytical models for the energy release rate of a substrate-adhesive-backing tape system and rigorously extending these formulations to large deformation cases. Specifically, the following four loading scenarios are investigated: (1) hanging a weight under the assumption of small deformation; (2) releasing the tape from the wall under small deformation; (3) hanging a weight under finite deformation; and (4) releasing the tape under finite deformation. As will be shown, the finite-deformation analytical solutions align closely with the numerical results from the finite element method (FEM) in the nonlinear regime, while the small-deformation theories severely deviate at high loads. Finally, FEM is utilized to extract the $J$-integrals at the crack tips of the two competing interfaces. By analyzing $J$-integrals across varying crack lengths, we explore the underlying fracture mechanics driving an alternating crack propagation pattern during the tape release process, and generate parametric maps that predict the required release force and the extent of unstable crack growth across various interfacial toughness combinations.

\section{Analytical solutions for the energy release rate and FEM verification}
\label{sec:analytical-solutions}

In this section, we first derive analytical solutions based on linearized elasticity and then extend these derivations to large deformations for an arbitrary hyperelastic solid.  In both cases, we assume that the backing and substrate layers are perfectly rigid, that both interface cracks are of the same length, and that the tape length is chosen to be much smaller than the shear-lag length \cite{wang2021lap} or the load transfer length \cite{hui2018mechanics,wang2020strength}, so that the shear stress remains uniform along the longitudinal direction $X_1$.  Since the out-of-plane width of the tape $w$ is much larger than the adhesive thickness and the adhesive is bonded to a rigid wall, plane strain deformation is assumed in the analysis.  Finally, the adhesive is assumed to be incompressible and hyperelastic. The key objective is to derive the energy release rate of the interface cracks for two use cases: (a) hanging a weight $W$ and (b) releasing with a force $F$.  In the following, the interface crack between the adhesive and backing layers will be referred to as the upper crack (where relevant quantities are denoted by a ``$+$'' subscript or superscript), whereas the interface crack between the adhesive and substrate layers is defined as the lower crack (denoted by a ``$-$'' subscript or superscript).

\subsection{Linear infinitesimal strain theory}

\subsubsection{Hanging weight}
\label{sec:linear-hanging}

When the strip is supporting a weight, the geometry and loading can be abstracted as shown in Fig.~\ref{fig:geometry}(b). Assuming that only the upper crack propagates, the energy release rate can be estimated by evaluating the change in strain energy associated with translating a material element of length $\mathrm{d} c$ and height $h_a$ from far ahead of the crack tip to far behind it. Far ahead of the crack tip, the material is in a state of simple shear with shear strain $\gamma$, and the corresponding strain energy stored in the element is $(\mu\gamma^2/2)wh_a \mathrm{d} c$, where $\mu$ is the small-strain shear modulus. The strain energy far behind the crack tip is approximately zero, hence the total energy release rate for the crack is
\begin{equation}
    G=\mu \gamma^2 h_a /2.
    \label{eq:linear-hanging-total-G}
\end{equation}
In the following analytical analysis, we consistently apply the approach of translating an imaginary material element for conciseness. Alternatively, a more traditional derivation of these results is computing the change in total strain energy per crack increment $\mathrm{d}c$, $\psi\mathrm{d}V=-\mu\gamma^2 wh_a\mathrm{d}c/2$ in this case, and the external work per crack increment, which is twice the change in strain energy under linear theory, to write the change in total potential energy $\mathrm{d}\Pi=-\psi\mathrm{d}V$ and derive $G$ through its definition $G=\mathrm{d}\Pi/\left(w \mathrm{d}c\right)$. 

Since the backing layer is rigid, most of the adhesive is in a state of simple shear, so an excellent estimate for $\gamma$ is $\gamma=\frac{1}{\mu}\frac{W}{wa}$. Substituting this relation into Eq.~\eqref{eq:linear-hanging-total-G} gives:
\begin{equation}
    G=\frac{1}{2\mu}\left(\frac{W}{wa}\right)^2 h_a.
\end{equation}
Note that in 2D plane strain, $W/w$ is simply the line load applied to the edge of the backing layer. If we conservatively assume that, when the upper crack propagates, most of the stored strain energy is released at the upper crack tip, and none of it flows to the lower crack tip, which we will justify later, then the energy release rate of the upper crack is: 
\begin{equation}
    G_w^+=\frac{1}{2\mu}\left(\frac{W}{wa}\right)^2 h_a.
    \label{eq:linear-hanging-upper-G}
\end{equation}
To determine the maximum weight $W_{\mathrm{max}}$ that can be supported by the tape, we assume that failure occurs if $G_w^+=\Gamma_{ba}$, where $\Gamma_{ba}$ is the interfacial toughness of the backing-adhesive interface. Using Eq.~\eqref{eq:linear-hanging-upper-G} and this criterion, we find
\begin{equation}
    W_{\mathrm{max}}=wa\sqrt{\frac{2\mu\Gamma_{ba}}{h_a}}.
    \label{eq:linear-hanging-critical}
\end{equation}

\subsubsection{Releasing from the wall}
\label{sec:linear-release}

To release the strip from the wall, the weight $W$ is removed and the exposed adhesive or the pull tab is stretched with a force $F$ alone, as illustrated in Fig.~\ref{fig:geometry}(c). Under load control, the energy release rate is determined by finding the change in potential energy by translating a material element of length $\mathrm{d}c$ and height $h_a$, from far in front of the crack tip to far behind the crack tip. The potential energy of material points far in front of the upper crack tip is zero since the adhesive there is stress-free, whereas the potential energy of a material point far behind the crack tip can be determined by the fact that the material is under plane-strain uniaxial tension, i.e., the strain $\varepsilon$ is given by
\begin{equation}
    \varepsilon=\frac{1}{E^*}\frac{F}{wh_a},
    \label{eq:linear-release-strain}
\end{equation}
where $E^*=4\mu$ is the plane-strain modulus for an incompressible solid. Therefore, the change in potential energy of a material element far behind the crack tip is 
\begin{equation}
    \mathrm{d}\Pi=-\frac{1}{8\mu}\left(\frac{F}{wh_a}\right)^2 wh_a\mathrm{d}c.
    \label{eq:linear-release-strain-energy}
\end{equation}
Similar to the hanging weight case, Eq.~\eqref{eq:linear-release-strain-energy} can also be derived through a traditional approach of finding the change in total strain energy and external work per crack increment, with the strain energy density being $\psi=F^2/\left(8\mu w^2h_a^2\right)$ and the change in total strain energy being $\psi\mathrm{d}V=F^2 \mathrm{d}c/\left(8\mu wh_a\right)$.

Using Eq.~\eqref{eq:linear-release-strain-energy} and the definition of energy release rate, the energy release rate of the upper crack $G_r^+$ is 
\begin{equation}
    G_r^+=\frac{1}{2}\frac{1}{8\mu}\left(\frac{F}{wh_a}\right)^2h_a.
    \label{eq:linear-release-upper-G}
\end{equation}
The factor of $1/2$ in Eq.~\eqref{eq:linear-release-upper-G} accounts for the fact that half of the strain energy flows to the lower crack, which is a reasonable approximation because of the symmetry of geometry and rigidity of the backing and substrate layers, and will be justified by the following numerical analysis in Section \ref{sec:fem}. Release is initiated when the energy release rate of the upper crack reaches the interfacial toughness $\Gamma_{ba}$. The force to initiate release or the critical release force $F_r$ is obtained by setting Eq.~\eqref{eq:linear-release-upper-G} to $\Gamma_{ba}$; this results in
\begin{equation}
    F_r=4wh_a\sqrt{\frac{\mu\Gamma_{ba}}{h_a}}.
    \label{eq:linear-release-critical}
\end{equation}


\subsubsection{Ratio of critical weight to critical release force}

Equations \eqref{eq:linear-hanging-critical} and \eqref{eq:linear-release-critical} allow us to compute the ratio:
\begin{equation}
    \frac{W_{\mathrm{max}}}{F_r}=\frac{\sqrt{2}a}{4h_a}.
    \label{eq:linear-critical-ratio}
\end{equation}
It is interesting to note that this ratio is \textit{independent} of material properties and depends only on a single geometrical parameter $a/h_a$. \textit{Since $a/h_a\gg 1$, the tape can support a very large weight but is still easy to remove.}

The linearized theory above suffers from several defects:
\begin{enumerate}
    \item Most adhesives have an elastic modulus less than $10^5$~Pa, hence they are highly extensible, as demonstrated by the 3M video of the release process. The assumption of small deformation is violated.
    \item The stress-strain relation of the adhesive is nonlinear.
\end{enumerate}

In addition, in the above analysis, we have assumed that only the upper crack propagates in both the hanging and release configurations. This assumption can be justified in the hanging case because the stresses and strains are higher at the upper adhesive interface (see upcoming numerical results). Therefore, unless the interfacial toughness of the adhesive-substrate interface $\Gamma_{as}$ is significantly lower than that of the adhesive-backing interface $\Gamma_{ba}$, the upper crack will initiate and propagate, while the lower crack remains stationary.

On the other hand, in the release configuration, the stress and strain states near the upper and lower crack tips are expected to be nearly identical. Consequently, it is possible for the lower crack to propagate first while the upper crack remains stationary, for example, if $\Gamma_{as}<\Gamma_{ba}$. In this case, the analysis is unchanged except that $\Gamma_{ba}$ in Eq.~\eqref{eq:linear-release-critical} is replaced by $\Gamma_{as}$. With this minor modification, Eq.~\eqref{eq:linear-critical-ratio} becomes:
\begin{equation}
    \frac{W_{\mathrm{max}}}{F_r}=\frac{\sqrt{2}a}{4h_a}\sqrt{\frac{\Gamma_{ba}}{\Gamma_{as}}}.
\end{equation}
Since $\Gamma_{ba}/\Gamma_{as}>1$, this ratio is slightly larger than in the case where only the upper crack is expected to propagate. 

It is interesting to note that, irrespective of which crack propagates first during release, the load is progressively transferred to the stationary crack. As a result, the advancing crack will eventually arrest, while the previously stationary crack might begin to propagate without increasing the load. Consequently, the release process is generally expected to involve alternating crack growth between the two interfaces, particularly when the two interfacial toughness values are comparable. However, in the determination of the release force above, we took the conservative estimate that release occurs once one of these cracks propagates. Later, in the numerical analysis, we will more precisely determine the release force, taking the alternating pattern into account.

\subsection{Finite deformation theory}
The following assumptions are made regarding the hyperelastic response of the adhesive. The adhesive is assumed to be isotropic and incompressible, so that $\mathrm{det}\mathbf{F}=1$, where $\mathbf{F}$ is the deformation gradient tensor. The strain energy density is therefore taken to have the general form
\begin{equation}
    \psi=\psi(I_1,I_2),
\end{equation}
where
\begin{equation}
    I_1=\mathrm{tr}\mathbf{C},\quad I_2=\frac{1}{2}\left[ \left(\mathrm{tr}\mathbf{C}\right)^2-\mathrm{tr}\left(\mathbf{C}^2\right)\right],
\end{equation}
and $\mathbf{C}=\mathbf{F}^{\mathrm{T}}\mathbf{F}$ is the right Cauchy–Green deformation tensor. However, under plane-strain conditions for an incompressible material, $I=I_1=I_2$. Consequently, there is no loss of generality in writing the strain energy density as 
\begin{equation}
    \psi=\Phi(I).
\end{equation}
We further assume that $\Phi$ is a monotonically increasing function satisfying $\Phi(3)=0$.

\subsubsection{Energy release rate for the hanging weight case}
\label{sec:finite-hanging}

The energy release rate for the upper crack is estimated using the strain energy density of the adhesive far ahead of the crack tip, $\psi_{\infty}$. The adhesive here is in a state of simple shear, with shear strain $\gamma_{\infty}$. 
\begin{equation}
    \psi_{\infty}=\Phi(I_{\infty})\Rightarrow G_w^+=h_a\Phi(I_{\infty}),
    \label{eq:finite-hanging-upper-G}
\end{equation}
where 
\begin{equation}
    I_{\infty}=3+\gamma_{\infty}^2.
\end{equation}
The nominal shear stress $S$ here is related to the hanging weight $W$ by
\begin{equation}
    S=\frac{W}{wa}=2\Phi'(I_{\infty})\gamma_{\infty}.
    \label{eq:finite-hanging-S}
\end{equation}
For a given weight and strain energy density function, we can solve for the shear strain $\gamma_{\infty}$ from Eq.~\eqref{eq:finite-hanging-S}, then use this shear strain to determine the energy release rate using Eq.~\eqref{eq:finite-hanging-upper-G}. Note that, just as in the small-strain derivation, Eq.~\eqref{eq:finite-hanging-upper-G} is an upper estimate, as we assume the energy release rate of the lower crack is zero, whereas in reality, some of the stored energy must flow to the lower crack. Later, we will show that the $J$-integral associated with the lower crack is not zero, although it is substantially smaller than that of the upper crack.

\subsubsection{Energy release rate for releasing}
\label{sec:finite-release}

Similar to our argument in the linear case, the energy release rate of the two cracks is determined by the change in potential energy associated with translating a material element from far ahead of the crack tip to the detached pull-tab where the force $F$ is applied. Since we expect that the two cracks have nearly the same energy release rate (see results of Section \ref{sec:fem} for justification), the energy release rate of the upper crack is half of this total energy, i.e.,
\begin{equation}
    G_r^+=\left[P\cdot(\lambda-1)-\Phi(I)\right]h_a/2,
    \label{eq:finite-release-upper-G-1}
\end{equation}
where $\lambda$ is the stretch ratio of the detached adhesive comprising the pull-tab, $I=\lambda^2+\lambda^{-2}+1$, and 
\begin{equation}
    P=F/(wh_a)
\end{equation}
is the axial nominal stress. As before, we assumed that the detached and significantly elongated part of the adhesive is well approximated by a state of uniaxial tension in plane strain, resulting in 
\begin{equation}
    P=\frac{F}{wh_a}=2\Phi'(I)\cdot \left(\lambda-\lambda^{-3}\right),
    \label{eq:finite-release-P}
\end{equation}
where the prime denotes the derivative with respect to $I$. Substituting Eq.~\eqref{eq:finite-release-P} into Eq.~\eqref{eq:finite-release-upper-G-1},
\begin{equation}
    G_r^+=\left[ 2\Phi'(I)\left(\lambda-\lambda^{-3}\right)(\lambda-1) -\Phi(I)\right]h_a/2.
    \label{eq:finite-release-upper-G-2}
\end{equation}
For a given force $F$ and a strain energy density function, we can solve for the stretch ratio $\lambda$ from Eq.~\eqref{eq:finite-release-P}, then use this stretch ratio to determine the energy release rate using Eq.~\eqref{eq:finite-release-upper-G-2}.

\subsubsection{Ratio of critical weight to critical release force}
As before, failure in the hanging weight case is said to occur when
\begin{equation}
    G_w^+=\Gamma_{ba}.
    \label{eq:finite-hanging-critical-equation}
\end{equation}
Thus, the maximum weight $W_{\mathrm{max}}$ that can be supported is obtained by solving for the critical shear strain using Eqs.~\eqref{eq:finite-hanging-upper-G} and \eqref{eq:finite-hanging-critical-equation}, then substituting this strain into Eq.~\eqref{eq:finite-hanging-S} to compute the maximum weight. For example, for a neo-Hookean solid, $2\Phi'(I)=\mu$, where $\mu$ is the small-strain shear modulus of the adhesive, the critical shear strain is obtained using Eqs.~\eqref{eq:finite-hanging-upper-G} and \eqref{eq:finite-hanging-critical-equation}, 
\begin{equation}
    G_w^+=\Gamma_{ba}\Leftrightarrow h_a\frac{\mu}{2}\gamma_{\infty}^2=\Gamma_{ba}\Rightarrow \gamma_{\infty}=\sqrt{\frac{2\Gamma_{ba}}{\mu h_a}}.
\end{equation}
The maximum weight is then (using Eq.~\eqref{eq:finite-hanging-S})
\begin{equation}
    W_{\mathrm{max}}=\mu\sqrt{\frac{2\Gamma_{ba}}{\mu h_a}}wa.
\end{equation}
We use the same criterion for release, i.e.,
\begin{equation}
    G_r^+=\Gamma_{ba},
    \label{eq:finite-release-critical-equation}
\end{equation}
where we have assumed that $\Gamma_{ba}<\Gamma_{as}$. The force $F_r$ required to release the tape is obtained by solving for the critical stretch ratio using Eqs.~\eqref{eq:finite-release-upper-G-2} and \eqref{eq:finite-release-critical-equation}, then substituting this stretch ratio to compute the critical release force. For the special case of a neo-Hookean solid, the critical stretch $\lambda_c$ is obtained by solving 
\begin{equation}
    \left(\lambda-\lambda^{-3}\right)(\lambda-1)-\frac{1}{2}\left(\lambda^2+\lambda^{-2}-2\right)=\frac{2\Gamma_{ba}}{h_a\mu}.
    \label{eq:finite-release-critical-stretch}
\end{equation}
The solution of Eq.~\eqref{eq:finite-release-critical-stretch} cannot be obtained in closed form, but a good approximation can be obtained by assuming the critical stretch is much larger than 1, hence Eq.~\eqref{eq:finite-release-critical-stretch} is approximately 
\begin{equation}
    \lambda(\lambda-1)-\frac{\lambda^2}{2}=\frac{2\Gamma_{ba}}{h_a\mu},
    \label{eq:finite-release-critical-stretch-2}
\end{equation}
where $\lambda^{-3}$, $\lambda^{-2}$, and the constant on the left-hand side are omitted as small quantities. The physically meaningful solution of Eq.~\eqref{eq:finite-release-critical-stretch-2} is
\begin{equation}
    \lambda_c=1+\sqrt{1+4\frac{\Gamma_{ba}}{h_a\mu}}.
\end{equation}
The critical force for release in this case is
\begin{equation}
    F_r=\mu \left(\lambda_c-\lambda_c^{-3}\right)wh_a\approx\mu\lambda_cwh_a=\mu w h_a\left(1+\sqrt{1+4\frac{\Gamma_{ba}}{h_a\mu}}\right).
    \label{eq:finite-release-critical}
\end{equation}
The ratio of maximum weight to release force is
\begin{equation}
 \frac{W_{\mathrm{max}}}{F_r}=\frac{a}{h_a}\left(\frac{\sqrt{\frac{2\Gamma_{ba}}{\mu h_a}}}{1+\sqrt{1+4\frac{\Gamma_{ba}}{h_a\mu}}}\right).
 \label{eq:finite-critical-ratio}
\end{equation}
Note that the ratio inside the parentheses is of order 1, and, since $a/h_a\gg 1$, \textit{the release force is much smaller than the maximum hanging weight}, like the prediction of the linearized theory given by Eq.~\eqref{eq:linear-critical-ratio}. Note that if $\Gamma_{ba}>\Gamma_{as}$, then the $\Gamma_{ba}$ in Eq.~\eqref{eq:finite-release-critical} and in the denominator of Eq.~\eqref{eq:finite-critical-ratio} should be replaced by $\Gamma_{as}$. Again, this assumes that only one of the cracks propagates when releasing, and we will more precisely determine $F_r$ through numerical results by investigating the alternating propagation pattern. 

For the general case, we note that most strain energy density functions can be written in the form:
\begin{equation}
    \Phi(I)=\mu\phi\left(I,\omega_1,\omega_2,\cdots,\omega_m\right),
    \label{eq:general-strain-energy-density}
\end{equation}
where $\phi$ is a monotonically increasing function of $I$, and $\omega_1,\omega_2,\cdots,\omega_m$ are dimensionless parameters governing the strain-hardening characteristics of the adhesive. A simple example is the Yeoh model \cite{yeoh1993some}, where 
\begin{equation}
    \Phi(I)=\sum_{k=1}^{m}C_k(I-3)^k=\frac{\mu}{2}\sum_{k=1}^m \omega_k (I-3)^k,
    \label{eq:yeoh}
\end{equation}
where $\omega_k=C_k/C_1$ and $C_1=\mu/2$. Another example is the Gent model \cite{gent1996new}, which accounts for finite extensibility, given by: 
\begin{equation}
    \Phi(I)=-\frac{\mu\omega}{2}\ln \left(1-\frac{I-3}{\omega}\right).
\end{equation}
Using the strain energy density function in Eq.~\eqref{eq:general-strain-energy-density} and following the same procedure above, it is easy to show that 
\begin{equation}
    \frac{W_{\mathrm{max}}}{F_r}=\frac{a}{h_a}g\left(\frac{\Gamma_{ba}}{h_a\mu},\omega_1,\cdots,\omega_m\right),
\end{equation}
where $g$ is a positive function of its dimensionless arguments. Hence the key result: \textit{the relationship } $\frac{W_{\mathrm{max}}}{F_r}\sim \frac{a}{h_a}\gg 1$ \textit{ holds for typical strain energy functions}. 

This result is important for design and can be understood physically: for the hanging mode, the weight is evenly distributed over a very large area of the adhesive, approximately $wa$, while during release, the peel force acts on a much smaller cross-sectional area, approximately $wh_a$, of the adhesive, creating a much larger stress concentration at the peel front.

\section{FEM verification of analytical energy release rate via \texorpdfstring{$J$}{J}-integral}
\label{sec:fem}

To support further analysis of the alternating crack propagation pattern, we first verify our assumptions and analytical solutions using Abaqus. All simulations are performed under a 2D plane strain assumption, consistent with the derivations above. The geometric setting and boundary conditions are consistent with Fig.~\ref{fig:geometry}(b)-(c). 

To properly resolve the singular fields, the mesh, consisting of 60,505 elements, is highly refined near the crack tips using structured CPE4H elements (a hybrid formulation featuring an independent hydrostatic pressure degree of freedom to ensure numerical stability in the nearly incompressible adhesive). A free-meshed transition region of CPE4H and CPE3H elements connects this tip refinement to a structured CPE4H far-field mesh. The $J$-integrals are computed over multiple circular contours, and the final analyzed values are averaged across the converged outer contours far from the crack tip. During loading, we also keep track of the strain in an element close to the left end of the adhesive or at the middle of the bonded adhesive to compute the stretch or shear strain.  

All inputs and results of the FEM are normalized. The only relevant length scale in the problem is the adhesive thickness $h_a$, since all other geometric dimensions are much larger than $h_a$. Accordingly, all lengths in the finite element simulations are normalized by $h_a$. Thus, the normalized length of the tape is $\bar{L}=L/h_a$, selected as 150. The upper and lower crack lengths are denoted by $c_+$ and $c_-$, respectively (see Fig.~\ref{fig:geometry}(b)-(c)), and their normalized counterparts are denoted by $\bar{c}_{\pm}$. These are parameterized by the distance between the two crack tips, $\bar{d}=\bar{c}_- - \bar{c}_+$, where either $\bar{c}_-$ or $\bar{c}_+$ is set to 50. All normalized quantities are topped with a bar.

According to our analytical solutions, the energy release rates $G_r^{\pm}$ and $G_w^{\pm}$ are proportional to $\mu h_a$ and thus can be normalized as 
\begin{equation}
    \bar{G}_r^{\pm}=G_r^{\pm}/(\mu h_a), \quad \bar{G}_w^{\pm}=G_w^{\pm}/(\mu h_a).
    \label{eq:normalized-G}
\end{equation}
In this and the following expressions, a bar denotes normalized quantities, and the ``$\pm$'' sign denotes the upper and lower cracks, respectively. The same normalization also applies to $J$-integrals, denoted as $\bar{J}^\pm$, and fracture toughnesses $\Gamma_{ba}$ and $\Gamma_{as}$. With the general form of strain energy density (Eq.~\eqref{eq:general-strain-energy-density}), applied loads can be normalized as
\begin{equation}
    \bar{W}=W/(\mu w a),\quad \bar{F}=F/(\mu w h_a).
\end{equation}
Dimensional considerations allow us to write
\begin{equation}
    \bar{G}_r^{\pm}=g_r^{\pm}\left(\bar{F},\bar{c}_+,\bar{c}_-,\bar{L}\right),\quad \bar{G}_w^{\pm}=g_w^{\pm}\left(\bar{W},\bar{c}_+,\bar{c}_-,\bar{L}\right).
    \label{eq:normalized-G-dimensional}
\end{equation}
Further simplification is possible by noting that $\bar{c}_+$, $\bar{c}_-$, and $\bar{L}$ are very large quantities compared to $h_a$ and $\bar{d}=\bar{c}_- - \bar{c}_+$; hence, Eq.~\eqref{eq:normalized-G-dimensional} can be simplified as:
\begin{equation}
    \bar{G}_r^{\pm}=g_r^{\pm}\left(\bar{F},\bar{d}\right),\quad \bar{G}_w^{\pm}=g_w^{\pm}\left(\bar{W},\bar{d}\right).
\end{equation}
Here we remind the reader that in the initial configuration, where we derived analytical solutions for both hanging and releasing modes, the cracks are of equal length, i.e., $\bar{d}=\bar{c}_- -\bar{c}_+=0$. Note that the functions $g_r^\pm$ and $g_w^{\pm}$ also depend on the hardening coefficients $\omega_1,\cdots,\omega_m$, which we have not included in their arguments to simplify notation.

In all the FEM simulations, the adhesive is modeled as a three-term Yeoh material (Eq.~\eqref{eq:yeoh}), with coefficients given by $\omega_1=1$, $\omega_2=-0.0474$, and $\omega_3=0.00332$. For $\bar{d}=0$, Eqs.~\eqref{eq:finite-hanging-upper-G} and \eqref{eq:finite-hanging-S} are now:
\begin{equation}
    \bar{G}_w^+\left(\bar{d}=0\right)=\frac{1}{2}\sum_{k=1}^3\omega_k\gamma_\infty^{2k},
    \label{eq:finite-hanging-upper-G-yeoh}
\end{equation}
where $\gamma_{\infty}$ is related to $\bar{W}$ by
\begin{equation}
    \bar{W}\left(\bar{d}=0\right)=\left[ 1+\sum_{k=2}^3k\omega_k\gamma_{\infty}^{2(k-1)}\right]\gamma_{\infty}.
    \label{eq:finite-hanging-force-strain-yeoh}
\end{equation}
Also, Eqs.~\eqref{eq:finite-release-upper-G-2} and \eqref{eq:finite-release-P} are now
\begin{equation}
    \bar{G}_r^+\left(\bar{d}=0\right)=\frac{1}{2}\left[1+\sum_{k=2}^3k\omega_k \left(\lambda^2+\lambda^{-2}-2\right)^{k-1}\right]\left(\lambda-\lambda^{-3}\right)(\lambda-1)-\frac{1}{4}\sum_{k=1}^{3}\omega_k\left(\lambda^2+\lambda^{-2}-2\right)^k,
    \label{eq:finite-release-upper-G-yeoh}
\end{equation}
\begin{equation}
    \bar{F}\left(\bar{d}=0\right)=\left[ 1+\sum_{k=2}^3 k\omega_k \left( \lambda^2+\lambda^{-2}-2 \right)^{k-1} \right]\left(\lambda-\lambda^{-3}\right).
    \label{eq:finite-release-force-strain-yeoh}
\end{equation}
These expressions can be used to solve for analytical strains and energy release rates.

\begin{figure}[t]
    \centering
    \includegraphics[width=0.80\linewidth]{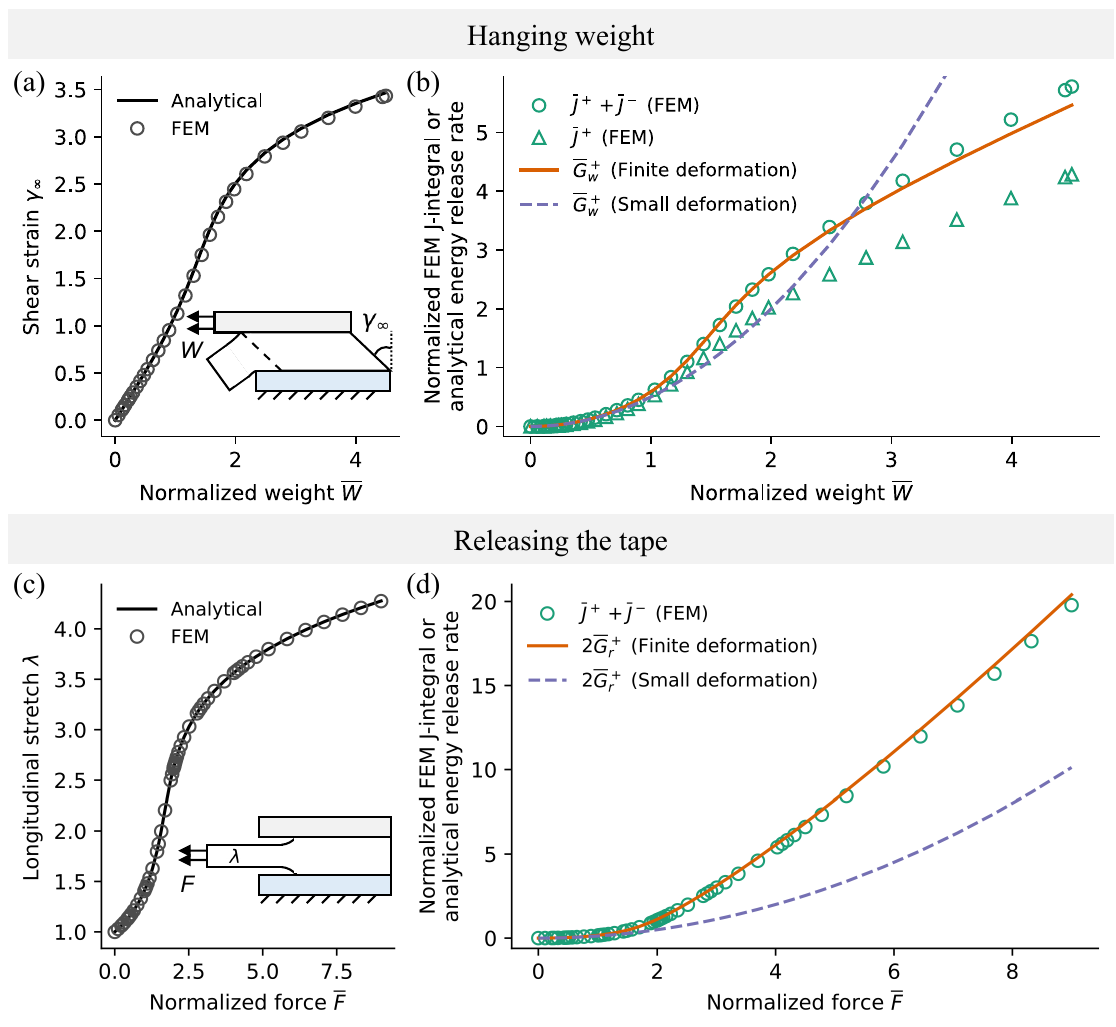}
    \caption{FEM results compared with analytical solutions for the weight hanging scenario (a) and (b), and for the tape releasing scenario (c) and (d). The analytical strains in (a) and (c) are solved numerically from Eqs.~\eqref{eq:finite-hanging-force-strain-yeoh} and \eqref{eq:finite-release-force-strain-yeoh}, respectively, and then substituted into Eqs.~\eqref{eq:finite-hanging-upper-G-yeoh} and \eqref{eq:finite-release-upper-G-yeoh} to compute the analytical energy release rate. The numerical shear strain in (a) is estimated by the deformation gradient component $F_{12}$. Results are for $\bar{d}=0$.}
    \label{fig:numerical-vs-analytical}
\end{figure}

Fig.~\ref{fig:numerical-vs-analytical} compares the numerical $J$-integral against the analytical energy release rates under both small- and finite-deformation assumptions for varying applied loads. At small deformations, the finite-deformation analytical models correctly reduce to the linear small-deformation limits, and both agree well with the numerical $J$-integral. Note that the analytical solution for $G_w^+$ is the total energy available for crack growth and is the upper estimate of the energy release rate of the upper crack; therefore, in Fig.~\ref{fig:numerical-vs-analytical}(b), $\bar{G}_w^+$ agrees with $\bar{J}^++\bar{J}^-$ and is higher than $\bar{J}^+$ alone. As the load increases and severe geometric and material nonlinearities begin to dominate, the small-deformation theory deviates significantly from the numerical results. Conversely, the finite-deformation theoretical solutions adhere closely to the $J$-integral, particularly in the releasing case (Fig.~\ref{fig:numerical-vs-analytical}(c)-(d)), maintaining a relative error of less than 10\% with respect to numerical $J$-integrals. Minor differences in the hanging weight case (Fig.~\ref{fig:numerical-vs-analytical}(a)-(b)) are because numerical results slightly deviate from the pure shear deformation in our analytical solutions. These results clearly validate our large-deformation analytical approach and demonstrate that the tape can bear much higher loads under the hanging weight case than under the releasing case, meaning it is much easier to release the tape than to fail the tape during normal usage.

\begin{figure}[t]
    \centering
    \includegraphics[width=0.85\linewidth]{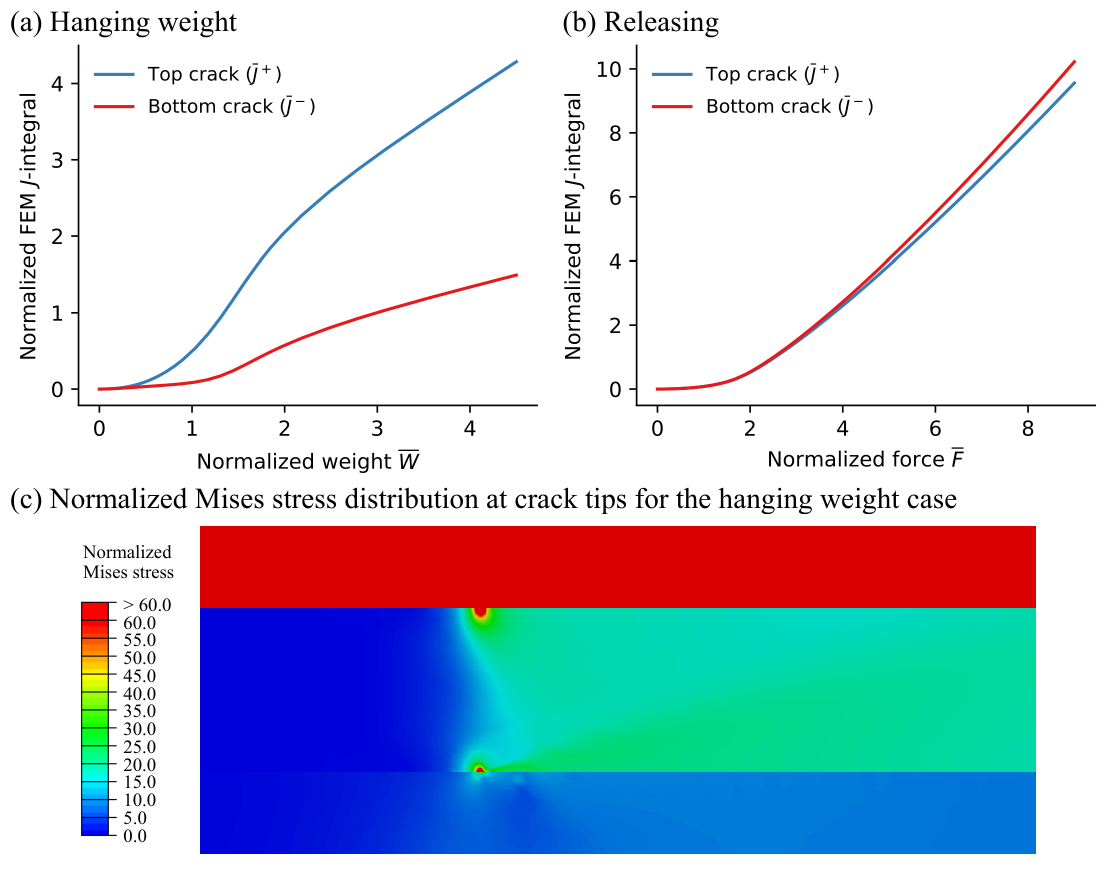}
    \caption{FEM $J$-integrals of the upper and lower crack tips, under both (a) the hanging weight and (b) releasing cases. Panel (c) illustrates the Mises stress distribution near the crack tips for the hanging weight case. Stress is normalized by the shear modulus.}
    \label{fig:top-vs-bot}
\end{figure}

The analytical solution for $G_w^+$ only tells how much energy can be released in total for the two cracks, but not how the value is distributed. Fig.~\ref{fig:top-vs-bot} shows numerical solutions for the $J$-integrals at both crack tips. For the releasing mode, $J$-integral values are very close with very minor deviations due to the slightly non-symmetric boundary conditions (the upper surface of the adhesive is traction-free); it is therefore reasonable to assume that both upper and lower cracks have the same energy release rate for $\bar{d}=0$, as we have assumed earlier in Section \ref{sec:linear-release} and \ref{sec:finite-release}.  Meanwhile, for the hanging mode, the upper crack has a much higher $J$ than the lower one, due to a higher stress concentration (Fig.~\ref{fig:top-vs-bot}(c)). Hence, our earlier assumption in Section \ref{sec:linear-hanging} and \ref{sec:finite-hanging} that our analytical solution of $G_w^+$ is a reasonable upper estimate is justified.  

Having rigorously validated the accuracy of the FEM framework against these analytical limits, we can now confidently deploy our numerical model to investigate the alternating crack propagation patterns observed on the two interfaces during the tape release process.

\section{Crack growth pattern and estimation on two interfaces when releasing}

\subsection{Crack propagation pattern}
\label{sec:crack-propagation-pattern}

\begin{figure}[t]
    \centering
    \includegraphics[width=1\linewidth]{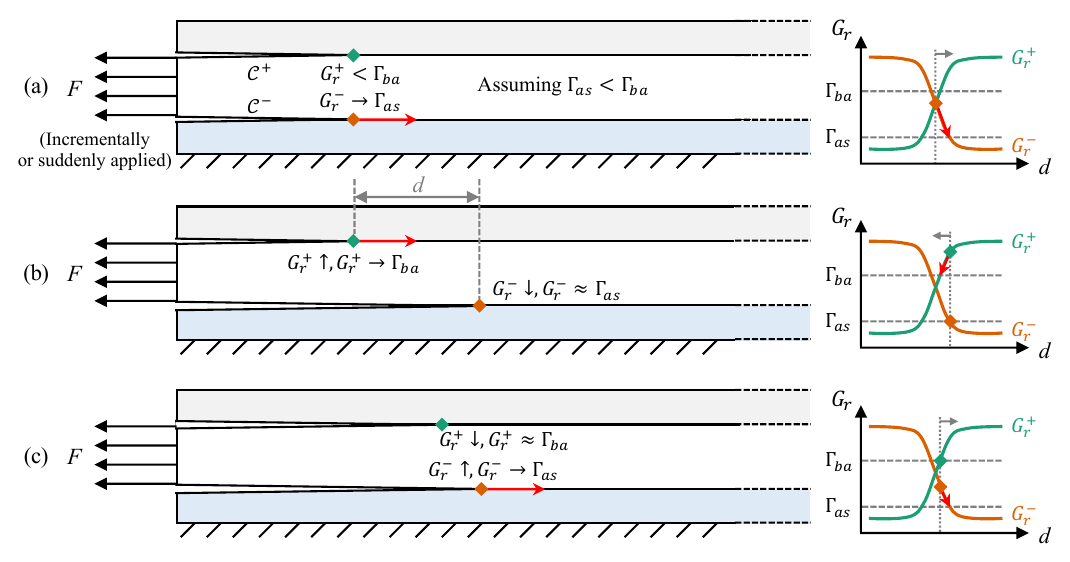}
    \caption{An idealized alternating crack propagation sequence on the two interfaces, with demonstrations of how $J$-integrals change during this process. This model assumes one crack grows and subsequently arrests when its driving force drops to the critical energy release rate, while the other crack remains stationary. This alternating mechanism is driven by geometric load transfer: as the propagating crack extends, its energy release rate decreases while the stationary crack’s energy release rate increases.}
    \label{fig:propagation-illustration}
\end{figure}

During the tape release process under a constant stretching force (see the user instruction video of 3M Command™ strips \cite{command_picture_strips_2026}), the adhesive undergoes severe finite deformation and gradually detaches from both the backing layer and the substrate. An idealized crack propagation sequence for this detachment is illustrated in Fig.~\ref{fig:propagation-illustration}. The adhesive generally exhibits different interfacial fracture toughnesses $\Gamma_{ba}$ (backing layer) and $\Gamma_{as}$ (substrate). In the following, we assume the substrate interface is the weaker of the two ($\Gamma_{as}<\Gamma_{ba}$), but the same argument applies to $\Gamma_{as}>\Gamma_{ba}$.

As one increases the force gradually from zero, eventually the energy release rate $G_r^-$ exceeds $\Gamma_{as}$, and the lower crack starts to grow at the weaker interface ($\mathcal{C}^-$). As it grows, meaning the distance between the two crack tips $d=c_--c_+$ increases, the energy release rate of the upper crack $G_r^+$ will increase while $G_r^-$ simultaneously decreases. Assuming a flat $R$-curve for both interfaces, $\mathcal{C}^-$ will propagate and subsequently arrest when its driving force $G_r^-$ drops due to the shifting boundary conditions and intersects the $R$-curve. Here we assume the increase in $G_r^+$ during this initial propagation is not large enough for it to reach $\Gamma_{ba}$, so the upper crack ($\mathcal{C}^+$) remains stationary. As the load continues to increase, the lower crack undergoes continuous stable growth to maintain $G_r^- = \Gamma_{as}$, which steadily increases $d$. When the load reaches a \textit{critical value} $F_r$, the energy release rate of the upper crack $G_r^+$ reaches its own critical threshold $\Gamma_{ba}$; at this point, both of the two cracks propagate without further increasing the load. This is the situation shown in the instruction video of 3M Command™ strips.

Instead of continuously increasing, the load can be applied incrementally or even rapidly in the real world, meaning the load may exceed the critical value $F_r$ after an increment. For simplicity and without loss of generality, we analyze a suddenly applied load $F$ above the critical load $F_r$, as demonstrated in Fig.~\ref{fig:propagation-illustration}, and \textit{assume that cracks propagate stably, ignoring dynamic effects}. We note that, when the suddenly applied load is equal to $F_r$, the following reasoning reduces to the above continuous-loading scenario. We will assume that only one crack propagates at any given time. In this scenario, $\mathcal{C}^-$ will propagate and subsequently arrest when $G_r^-$ drops to $\Gamma_{as}$ (Fig.~\ref{fig:propagation-illustration}(a)). As $d$ increases, $G_r^+$ goes above $\Gamma_{ba}$ and $\mathcal{C}^+$ will then propagate until it arrests (Fig.~\ref{fig:propagation-illustration}(b)), which in turn alters the geometry ($d$ decreases) to elevate $G_r^-$ again, re-initiating $\mathcal{C}^-$ as $G_r^-$ increases above $\Gamma_{as}$ after the load transfer (Fig.~\ref{fig:propagation-illustration}(c)). This process repeats recursively without further increasing the load, resulting in an alternating crack propagation pattern. A similar phenomenon is the stick-slip crack growth pattern \cite{zheng2024unique}, an alternating propagation-arresting process of a standalone crack, while the alternating pattern we discuss here is driven by the interaction of two cracks. 

The sequence will not occur if the suddenly applied load is below $F_r$, since the propagation of either $\mathcal{C}^-$ or $\mathcal{C}^+$ fails to drive the opposing crack’s energy release rate above its critical threshold, at which point the external force must be increased to drive further detachment. Formally, to trigger the alternating propagation given $\Gamma_{as}<\Gamma_{ba}$, we require a sufficiently high $F$ such that $G_r^-(d=0,F)\geq\Gamma_{as}$, followed by $G_r^+(d^-,F)\geq\Gamma_{ba}$ (where $d^-$ is the solution to $G_r^-(d^-,F)=\Gamma_{as}$), and subsequently $G_r^-(d^+,F)\geq\Gamma_{as}$ (where $d^+$ is the solution to $G_r^+(d^+,F)=\Gamma_{ba}$). 

It should be noted that this alternating model strictly assumes isolated, sequential propagation. The simultaneous unstable propagation of both cracks, which would involve complex dynamic effects such as elastic wave propagation and precise transient $R$-curve coupling, is beyond the scope of this static analysis.

\subsection{Crack propagation estimation}

\begin{figure}[t]
    \centering
    \includegraphics[width=1\linewidth, trim={1cm 10.5cm 4.5cm 0cm}, clip]{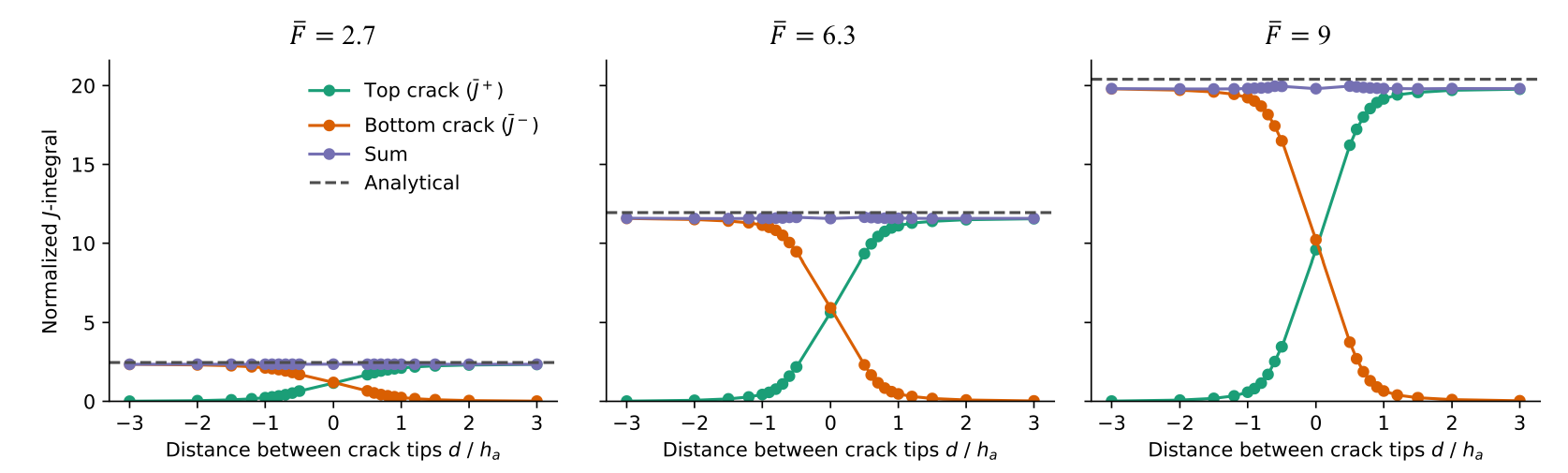}
    \caption{Relation between the $J$-integrals at the two crack tips and the normalized longitudinal distance between them ($\bar{d}=d/h_a$).}
    \label{fig:j-versus-d}
\end{figure}

By utilizing the finite-deformation FEM framework with the Yeoh model, we can map the exact dependence of the normalized geometric driving forces $\bar{J}^+$ and $\bar{J}^-$ on the normalized longitudinal distance between the two crack tips, $\bar{d}=\bar{c}_--\bar{c}_+$ (Fig.~\ref{fig:j-versus-d}), as well as on the normalized applied load $\bar{F}$. Assuming flat $R$-curves for both interfaces, these calibrated $J$-integral functions allow us to analytically or algorithmically predict the evolution of both crack tips and the distance between them during the alternating propagation process under a given load. 

In the following, we present two approaches to analyze the alternating propagation process.

\noindent \textbf{Approach 1: Analytically solve for the critical release load}

Analytically, we can solve for the critical load $\bar{F}_r$ needed to trigger the alternating propagation, as noted in Section \ref{sec:crack-propagation-pattern}, by finding the minimal $\bar{F}$ that satisfies all of $\bar{J}^-\left(\bar{d}=0,\bar{F}\right)\geq\bar{\Gamma}_{as}$, $\bar{J}^+\left(\bar{d}^-,\bar{F}\right)\geq\bar{\Gamma}_{ba}$ (where $\bar{d}^-$ is the solution to $\bar{J}^-\left(\bar{d}^-,\bar{F}\right)=\bar{\Gamma}_{as}$), and $\bar{J}^-\left(\bar{d}^+,\bar{F}\right)\geq\bar{\Gamma}_{as}$ (where $\bar{d}^+$ is the solution to $\bar{J}^+\left(\bar{d}^+,\bar{F}\right)=\bar{\Gamma}_{ba}$), when $\bar{\Gamma}_{as}<\bar{\Gamma}_{ba}$. Similar requirements can be easily inferred for $\bar{\Gamma}_{as}>\bar{\Gamma}_{ba}$. $\bar{J}^+$ and $\bar{J}^-$ can be calibrated as explicit functions of $\bar{d}$ and $\bar{F}$ using data shown in Fig.~\ref{fig:j-versus-d} ($\bar{F}=9$) and Fig.~\ref{fig:top-vs-bot}(b):
\begin{equation}
    \bar{J}^{+}(\bar{d}, \bar{F})=0.106\left(-0.013 \bar{F}^{3}+0.253 \bar{F}^{2}-0.169 \bar{F}\right)\left[9.930 \tanh \left(1.616 \bar{d}\right)+9.855\right] 
\end{equation}
\begin{equation}
\bar{J}^{-}(\bar{d}, \bar{F})=0.099\left(-0.013 \bar{F}^{3}+0.264 \bar{F}^{2}-0.181 \bar{F}\right)\left[-9.923 \tanh \left(1.619 \bar{d}\right)+9.989\right]
\end{equation}
Note that these expressions are calibrated for the specific Yeoh model and are normalized to be independent of the shear modulus and geometric dimensions. 

\begin{figure}[t]
    \centering
    \includegraphics[width=0.5\linewidth]{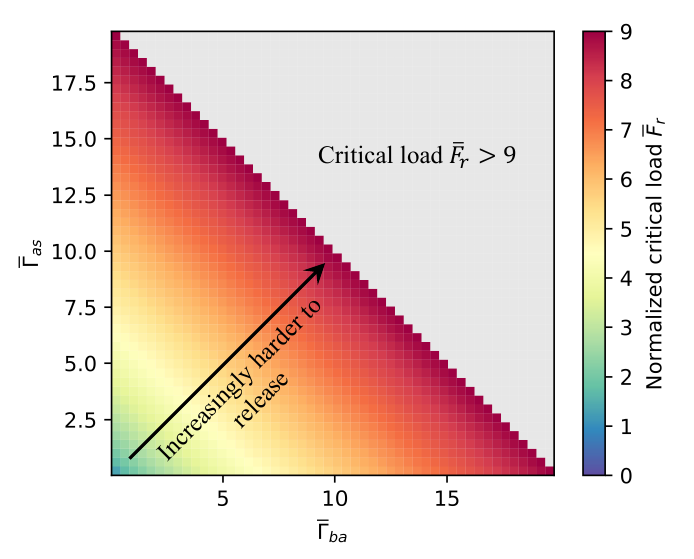}
    \caption{Critical release load given different combinations of $\bar{\Gamma}_{ba}$ and $\bar{\Gamma}_{as}$. Values of $\bar{F}_r$ greater than 9 are masked in gray. Our data only covers $\bar{F}_r\leq 9$, and extrapolated results may be unreliable.}
    \label{fig:critical-load}
\end{figure}

Using this approach, critical load values $\bar{F}_r$ for this specific Yeoh model are shown in Fig.~\ref{fig:critical-load}. Increasing either interface toughness will make it harder to trigger the alternating propagation and thereby release the tape. Since $\bar{J}^+$ and $\bar{J}^-$ are almost symmetric (Fig.~\ref{fig:j-versus-d}), the critical load is nearly symmetric about the line $\bar{\Gamma}_{as}=\bar{\Gamma}_{ba}$. 

\noindent \textbf{Approach 2: Algorithmically simulate the alternating crack propagation}

Algorithmically, we can exactly replicate the process described in Section \ref{sec:crack-propagation-pattern} to compute the amount of crack propagation iteratively, until the process halts or a tape of a certain length ($100h_a$) is fully released. We can identify not only the critical load but also the details of the crack extension iterations. The methodology for this idealized, iterative simulation is detailed in \ref{sec:algorithm}. We also note that identical results can be obtained by processing $\bar{d}^-$ and $\bar{d}^+$ solved in the analytical Approach 1. 

\begin{figure}[t]
    \centering
    \includegraphics[width=1\linewidth, trim={1cm 1cm 4.5cm 9cm}, clip]{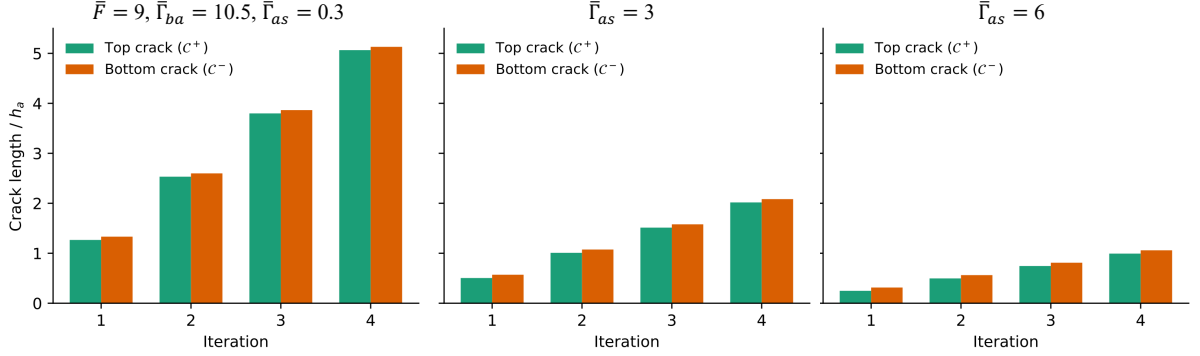}
    \caption{The estimated crack propagation pattern, given $\bar{F}=9$, $\bar{\Gamma}_{ba}=10.5$, and varying $\bar{\Gamma}_{as}$.}
    \label{fig:estimated-propagation}
\end{figure}

From the iterative algorithm, Fig.~\ref{fig:estimated-propagation} illustrates the crack extension over the first four iterations under a constant peel force $\bar{F}=9$, assuming $\bar{\Gamma}_{ba}=10.5$ and varying the substrate toughness ($\bar{\Gamma}_{as}=0.3, 3, 6$). Notably, by the second iteration, the system achieves a steady alternating propagation pattern where the incremental crack growth per iteration becomes constant.

Building upon this iterative algorithm, we can conduct parametric sweeps over the design space ($\bar{\Gamma}_{ba}$, $\bar{\Gamma}_{as}$, and $\bar{F}$) to quantify macroscopic performance metrics: the total number of iterations required for the tape to completely detach, and the maximum normalized separation distance $\bar{d}$ between the cracks during steady-state peeling.

\begin{figure}[t]
    \centering
    \includegraphics[width=1\linewidth, trim={4.5cm 10.3cm 8.5cm 0cm}, clip]{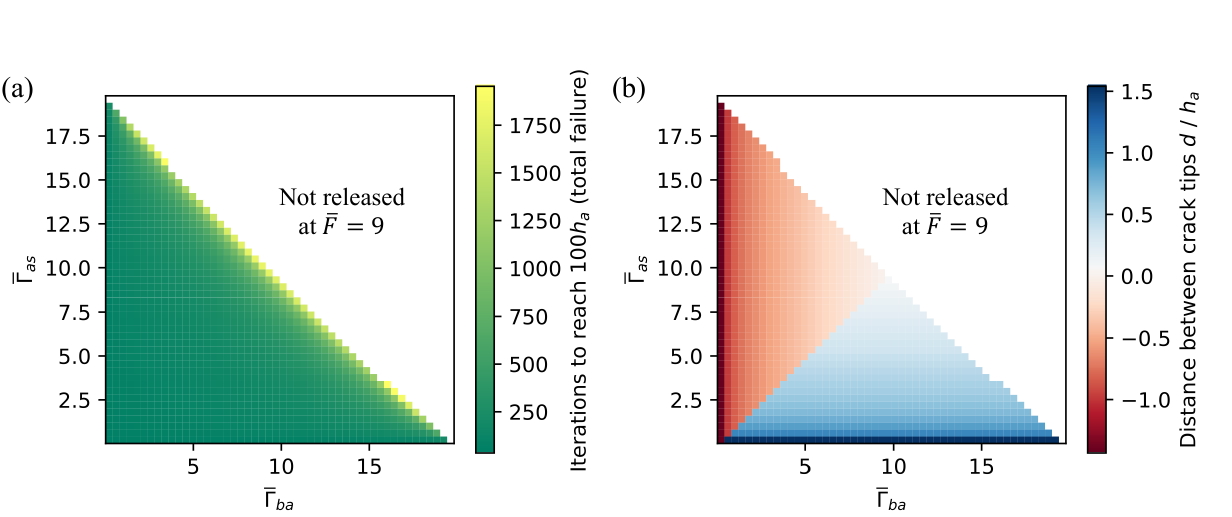}
    \caption{Given a constant release force $\bar{F}=9$ across varying combinations of $\bar{\Gamma}_{ba}$ and $\bar{\Gamma}_{as}$: (a) the total iterations required to completely detach a tape of length $100h_a$, and (b) the resulting steady-state normalized distance $\bar{d}$ between the crack tips, after the crack on the weaker interface propagates.}
    \label{fig:sweep-gc1-gc2}
\end{figure}

Sweeping across a wide range of $\bar{\Gamma}_{ba}$ and $\bar{\Gamma}_{as}$ for a fixed load ($\bar{F}=9$), we calculate the iterations required to fully release the remaining tape length of $100h_a$. Fig.~\ref{fig:sweep-gc1-gc2} summarizes these results via heat maps. The colored regions designate parametric combinations where the applied force is sufficient to successfully release the tape. Notice that the colored region for successful release matches the region in Fig.~\ref{fig:critical-load} where $\bar{F}_r\leq 9$, cross-validating the two approaches. A higher iteration count implies shorter individual crack jumps, rendering the propagation process more susceptible to arrest by local interfacial inhomogeneities. Meanwhile, a larger tip separation $\bar{d}$ correlates with extensive unstable propagation on the weaker interface. For instance, when one interface is exceptionally weak ($\bar{\Gamma}_{ba}$ or $\bar{\Gamma}_{as}\rightarrow 0$), the separation distance reaches a theoretical maximum. Conversely, the blank (white) regions indicate safe operational zones (not released under the given load); here, the critical fracture toughness of at least one interface strictly exceeds the maximum available energy release rate with varying geometric settings, permanently arresting the detachment. 

\begin{figure}[t]
    \centering
    \includegraphics[width=1\linewidth, trim={4.5cm 1.5cm 8.5cm 10cm}, clip]{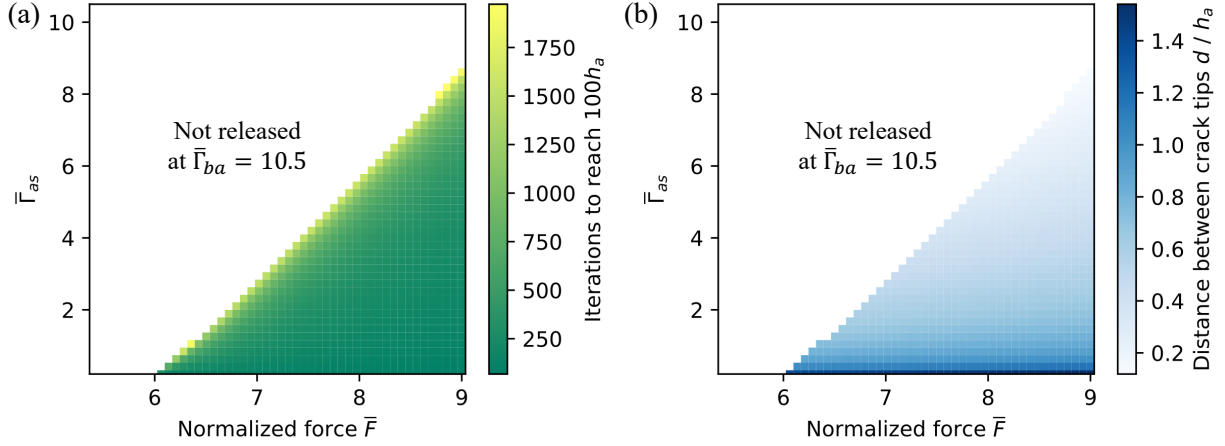}
    \caption{Given a fixed backing toughness $\bar{\Gamma}_{ba}=10.5$ across varying combinations of $\bar{\Gamma}_{as}$ and $\bar{F}$: (a) iterations needed for complete detachment, and (b) the resulting normalized distance between the crack tips.}
    \label{fig:sweep-gc2-f}
\end{figure}

In commercial products such as 3M Command™ strips, the backing layer and adhesive are predetermined, making $\bar{\Gamma}_{ba}$ fixed. The attached surface, however, can vary from application to application. By fixing $\bar{\Gamma}_{ba}$ and sweeping over the substrate toughness $\bar{\Gamma}_{as}$ and applied force $\bar{F}$ (Fig.~\ref{fig:sweep-gc2-f}), we construct a holistic failure envelope. This envelope dictates the exact required release force for any given wall substrate. As shown in Fig.~\ref{fig:critical-load}, as the substrate adhesion ($\bar{\Gamma}_{as}$) strengthens, the required release force $\bar{F}_r$ strictly increases. Meanwhile, the alternating crack growth proceeds with significantly less unstable propagation per iteration.

\section{Conclusion}

This study rigorously investigated the fracture mechanics of a hyperelastic tape system under finite deformation. By extending classical small-deformation analytical models, explicit governing equations for the energy release rate were derived for both the hanging weight and tape release scenarios for incompressible materials under the plane-strain assumption. These analytical solutions allow us to explain why releasing the tape requires much lower force than breaking the tape under the hanging scenario. 

These analytical limits were validated against numerical $J$-integrals extracted from a highly refined, hybrid-element FEM framework. While linear elastic fracture mechanics (LEFM) theories severely deviated from physical reality at higher loads, the finite-deformation formulations maintained excellent accuracy, capturing the geometric and material nonlinearities with less than 10\% relative error. Note that we apply a tape length that is much smaller than the load transfer length of an elastic backing layer \cite{hui2018mechanics}, meaning the shear stress exerted by the backing layer stays uniform along the longitudinal direction. With a much longer tape or a much softer backing layer, the shear stress becomes non-uniform, and the analytical solutions become invalid.

Finally, by deploying this validated FEM model, an alternating crack propagation mechanism was identified, analytically solved, and algorithmically simulated. By analyzing the geometric load transfer and compliance shifts between the competing backing and substrate interfaces, we obtain the critical load needed to release the tape for different interface toughnesses, and the iterative propagation algorithm can capture the details of the alternating crack propagation behavior. These results provide a holistic predictive tool for estimating the required release force, the extent of unstable crack propagation, and the ultimate operational limits of the tape system across various real-world substrates. The analysis in this study can bolster the design of the stretch-release mechanism in other applications to ensure convenience and reliability. 


\section*{Statements and Declarations}

\noindent \textbf{Data availability.} Data and code are available on GitHub (\url{https://github.com/LuoXueling/3m_command_strip_analysis}). 

\noindent \textbf{Acknowledgments.} C.Y. Hui acknowledges the support of the National Science Foundation under Grant No. CMMI-1903308. C.Y. Hui acknowledges Prof. Rong Long for bringing this problem to his attention. The derivation of the linearized theory was partially based on material presented in Prof. Long's fracture mechanics class.

\noindent \textbf{Competing Interests.} The authors declare no competing interests.

\newpage
\appendix

\section{Algorithm for estimating crack propagation pattern}
\label{sec:algorithm}
{\setstretch{1.1}
\vspace{0.3cm}
\begin{algorithm}[H]
\SetAlgoLined
\DontPrintSemicolon
\vspace{0.1cm}
\KwIn{Normalized distance--$\bar{J}$ data for $\mathcal{C}^+$ and $\mathcal{C}^-$, $\bar{\Gamma}_{ba}$, $\bar{\Gamma}_{as}$, max iterations $N_{\max}$}
\KwOut{Per-iteration propagation and cumulative propagation for both cracks}
\vspace{0.1cm}

Build natural cubic splines $\bar{J}^+(\bar{d})$, $\bar{J}^-(\bar{d})$\;
Set $\bar{d} \leftarrow 0$\;
Set $\bar{c}_+ \leftarrow 0$, $\bar{c}_- \leftarrow 0$\;

\For{$k=1$ \KwTo $N_{\max}$}{
    $\Delta\bar{d}^+ \leftarrow 0$, $\Delta\bar{d}^- \leftarrow 0$\;

    \tcp{Determine evaluation sequence based on weaker interface}
    Set $Seq \leftarrow [1, 2]$ \textbf{if} $\bar{\Gamma}_{ba} \le \bar{\Gamma}_{as}$ \textbf{else} $[2, 1]$\;

    \For{$i \in Seq$}{
        \eIf{$i = 1$}{
            \If{$\bar{J}^+(\bar{d}) > \bar{\Gamma}_{ba}$}{
            Find $\bar{d}_1$ such that $\bar{J}^+(\bar{d}_1) = \bar{\Gamma}_{ba}$ with $\bar{d}_1 \le \bar{d}$\;
                $\Delta\bar{d}^+ \leftarrow \bar{d} - \bar{d}_1$; $\bar{c}_+ \leftarrow \bar{c}_+ + \Delta\bar{d}^+$; $\bar{d} \leftarrow \bar{d}_1$\;
            }
        }{
            \If{$\bar{J}^-(\bar{d}) > \bar{\Gamma}_{as}$}{
            Find $\bar{d}_2$ such that $\bar{J}^-(\bar{d}_2) = \bar{\Gamma}_{as}$ with $\bar{d}_2 \ge \bar{d}$\;
                $\Delta\bar{d}^- \leftarrow \bar{d}_2 - \bar{d}$; $\bar{c}_- \leftarrow \bar{c}_- + \Delta\bar{d}^-$; $\bar{d} \leftarrow \bar{d}_2$\;
            }
        }
    }

    Record $(k, \Delta\bar{d}^+, \Delta\bar{d}^-, \bar{c}_+, \bar{c}_-)$\;

    \If{$\bar{J}^+(\bar{d}) \le \bar{\Gamma}_{ba}$ \textbf{and} $\bar{J}^-(\bar{d}) \le \bar{\Gamma}_{as}$}{
        \textbf{break}\;
    }
}
\caption{Iterative crack-propagation estimation using spline-interpolated $\bar{J}^{\pm}(\bar{d})$}
\label{algo:iterative-simulation}
\end{algorithm}
}



\begin{thebibliography}{10}
\expandafter\ifx\csname url\endcsname\relax
  \def\url#1{\texttt{#1}}\fi
\expandafter\ifx\csname urlprefix\endcsname\relax\def\urlprefix{URL }\fi
\expandafter\ifx\csname href\endcsname\relax
  \def\href#1#2{#2} \def\path#1{#1}\fi

\bibitem{command_picture_strips_2026}
{3M Company}, \href{https://www.command.com/3M/en_US/command/how-to-use/picture-hanging-strips/}{How to use command™ strips for hanging pictures} (2026).
\newline\urlprefix\url{https://www.command.com/3M/en_US/command/how-to-use/picture-hanging-strips/}

\bibitem{command_products_2026}
{3M Company}, \href{https://www.command.com/3M/en_US/command/how-to-use/}{How to use command™ strips and hooks} (2026).
\newline\urlprefix\url{https://www.command.com/3M/en_US/command/how-to-use/}

\bibitem{kendall1975thin}
K.~Kendall, Thin-film peeling-the elastic term, Journal of Physics D: Applied Physics 8~(13) (1975) 1449--1452.

\bibitem{kendall1975crack}
K.~Kendall, Crack propagation in lap shear joints, Journal of Physics D: Applied Physics 8~(5) (1975) 512--522.

\bibitem{hui2018mechanics}
C.-Y. Hui, Z.~Liu, H.~Minsky, C.~Creton, M.~Ciccotti, Mechanics of an adhesive tape in a zero degree peel test: effect of large deformation and material nonlinearity, Soft Matter 14~(47) (2018) 9681--9692.

\bibitem{kaelble1960theory}
D.~H. Kaelble, Theory and analysis of peel adhesion: bond stresses and distributions, Transactions of the Society of Rheology 4~(1) (1960) 45--73.

\bibitem{liu2019mechanics}
Z.~Liu, H.~Minsky, C.~Creton, M.~Ciccotti, C.-Y. Hui, Mechanics of zero degree peel test on a tape—effects of large deformation, material nonlinearity, and finite bond length, Extreme Mechanics Letters 32 (2019) 100518.

\bibitem{wang2020strength}
Y.~Wang, X.~Yang, G.~Nian, Z.~Suo, Strength and toughness of adhesion of soft materials measured in lap shear, Journal of the Mechanics and Physics of Solids 143 (2020) 103988.

\bibitem{wang2021lap}
Y.~Wang, G.~Nian, X.~Yang, Z.~Suo, Lap shear of a soft and elastic adhesive, Mechanics of Materials 158 (2021) 103845.

\bibitem{horgan2017poynting}
C.~Horgan, J.~Murphy, Poynting and reverse poynting effects in soft materials, Soft matter 13~(28) (2017) 4916--4923.

\bibitem{yeoh1993some}
O.~H. Yeoh, Some forms of the strain energy function for rubber, Rubber Chemistry and technology 66~(5) (1993) 754--771.

\bibitem{gent1996new}
A.~N. Gent, A new constitutive relation for rubber, Rubber chemistry and technology 69~(1) (1996) 59--61.

\bibitem{zheng2024unique}
Y.~Zheng, Y.~Wang, F.~Tian, T.~Nakajima, C.-Y. Hui, J.~P. Gong, Unique stick--slip crack dynamics of double-network hydrogels under pure-shear loading, Proceedings of the National Academy of Sciences 121~(30) (2024) e2322437121.

\end{thebibliography}
\end{document}